# Effect of Pre-inflation Conditions on Scalar and Tensor Perturbations in Inflation

Shiro Hirai *

Department of Digital Games, Faculty of Information Science and Arts,
Osaka Electro-Communication University
Neyagawa, Osaka 572-8530, Japan
*E-mail*: hirai@isc.osakac.ac.jp

Abstract

The effect of the initial conditions in inflation on scalar and tensor perturbations is investigated. Formulae for the power spectra of gravitational waves and curvature perturbations for any initial conditions in inflation are derived, and the ratio of scalar to tensor perturbations and spectral index are calculated. The formulation is applied to some simplified pre-inflationary cosmological models, and the differences of the ratio and spectral index are investigated with respect to two matching conditions. In addition, the present power spectrum of gravitational waves is derived. The proposed formulation is preliminarily shown to be a possible test of the appropriateness of a given pre-inflationary model.

## 1. Introduction

One of the most interesting problems in inflationary cosmology is that of the scalar and tensor perturbations in inflation. These perturbations are considered to have become the seeds of the present galaxies, and this area has been investigated extensively based on measurements of cosmic microwave background anisotropy. The power spectrum of gravitational waves has been calculated by some methods and with some models.[1) ] Recently, Giovannini[2)] investigated the production of relic gravitons in quintessential inflationary models, and showed that the production of gravitons is greatly enhanced at high gravitational wave frequency ($10^{-3}$ Hz < $\nu$ < $10^{10}$ Hz) due to the stiff phase after inflation. In the previous report,[3)] the effect of a non-zero cosmological constant on background gravitational waves was investigated using a simple cosmological model with the cosmological constant. The power spectrum of the curvature perturbations produced during inflation is also very important, and is believed to be fixed at the horizon crossing in inflation and to change little until the perturbation reenters the Hubble horizon in the radiation-dominated period (matter-dominated period). The first- and second-order corrections to this power spectrum were obtained assuming a slow-roll expansion.[4,5)] Contrary to the above property of the curvature perturbations a model in which inflation is temporarily suspended has been proposed and a large amplification of the curvature perturbation relative to its value at horizon crossing has been found.[6)]

It can be inferred from string/M-theory that several physical circumstances may have existed before inflation. The importance of the initial condition of inflation[7)] and the trans-Planckian problem of inflationary cosmology[8)] have therefore been discussed at great length. In a previous study,[9)] the effect of the initial condition in inflation on the power spectrum of the curvature perturbations was examined from another point of view. Based on the physical conditions before inflation, the possibility exists that the initial state of the scalar perturbations is not only the Bunch-Davies state, but also a more general state – a squeezed state, as occurs between two phases, such as between the inflation and radiation-dominated epochs.[10,11)] Based on this consideration, a formula for the power spectrum of the curvature perturbations given any initial conditions in inflation was derived. Here, this concept is extended to gravitational waves having any initial condition in inflation, leading to a new formulation of the tensor-scalar ratio and spectral index.

The power spectrum of gravitational waves is commonly derived using the Bunch-Davies vacuum as the initial condition of inflation.[1-3] The power spectrum of gravitational waves in this study is derived for any initial conditions (a squeezed state) in inflation by multiplying the familiar formation by a factor indicating the contribution of the initial condition. This form is very similar to the case for curvature perturbations, and a similar formation has been derived by Hawang.[12] The derived formula is applied here to a range of cases and models to clarify its physical meaning. Using the power spectrum of the curvature perturbations for any initial conditions in inflation as derived in the previous report, a formula for the ratio of the power spectrum of the gravitational waves to that of the curvature perturbations is derived. The influence of arbitrary initial conditions in inflation on spectral indices is examined for both curvature perturbations and gravitational waves.

The physical meanings of the derived formulae are clarified through examination of a number of pre-inflationary cosmological models and calculation of the ratio and spectral index for each case. The matching conditions of the scalar perturbations, which appear to be essential and have been discussed in relation to inflation[13] and an ekpyrotic scenario,[14] are also examined here. Calculations are made assuming two matching conditions; one in which the gauge potential and its first $\eta$ -derivative are continuous at the transition point, and one in which the transition happens on a hyper-surface of constant energy as proposed by Deruelle and Mukhanov,[13)] and the differences between the two matching conditions are investigated. In addition, the present power spectrum of gravitational waves for any initial conditions in inflation is derived as a simple form, and is calculated and compared to the regular power spectrum assuming a simple cosmological model.

In section 2, formulae for the power spectra of the tensor and scalar perturbations for any initial conditions in inflation are derived, and the tensor-to-scalar ratio and spectral index are obtained. In section 3, some simplified cosmological models used to demonstrated the present formulation are outlined. In section 4, the power spectra of the tensor and scalar perturbations, the tensor-to-scalar ratio and the spectral indices are calculated using the derived formula and a number of pre-inflationary cosmological models. The effect of the initial conditions are estimated, and the difference between the two matching conditions are investigated. In section 5, the present power spectrum of gravitational waves for any initial conditions in inflation is derived, and in section 6, the results obtained in the present study are discussed.

## 2. Tensor and scalar perturbations

In this section, formulae for the power spectra of the tensor and scalar perturbations for any initial conditions (a squeezed initial state) in inflation are derived, and using these formulae the tensor-to-scalar ratio and spectral indices are obtained.

2.1 Tensor perturbations

A spatially flat Friedman-Robertson-Walker (FRW) universe is considered for the background spectrum, as described by the metric

$$ds^2 = a^2(\eta)(-d\eta^2 + dx^2 + dy^2 + dz^2), \tag{2.1}$$

where $\eta$ is the conformal time. Gravitational waves are deviation from the geometry of this background metric, as given by

$$g_{ij} = a^2(\eta)(\delta_{ij} + h_{ij}), \tag{2.2}$$

where $h_{ij}$ describes a small deviation from $\delta_{ij}$. Under the transverse traceless gauge, the action of the gravitational waves in the linear approximation is given by

$$S = \frac{1}{2}\int d^4x \{(\frac{\partial h}{\partial\eta})^2 - (\nabla h)^2 + \frac{a''}{a} h^2\}, \tag{2.3}$$

where $h$ is the transverse traceless part of the deviation $h_{ij}$ and represents the two independent polarization states of the wave $(h_+, h_\times)$.[15)] The primes denote the derivative with respect to conformal time $\eta$. As is well known, the action of gravitational waves in the linear approximation is in the same form as that for a real massless, minimally coupled scalar field.[15)] The field $h(\eta, \boldsymbol{x})$, can be expanded in terms of the annihilation and creation operators $\boldsymbol{a_k}$ and $\boldsymbol{a_{-k}}^\dagger$, i.e.,

$$h(\eta, x) = \frac{1}{(2\pi)^{3/2}a}\int d^3k\{ v_k(\eta)\, \boldsymbol{a_k} + v_k^*(\eta)\, a_{-k}{}^\dagger\}e^{-ikx}, \tag{2.4}$$

and $v_k(\eta)$ is the solution of

$$\frac{d^2 v_k}{d\eta^2} + (k^2 - \frac{1}{a}\frac{d^2 a}{d\eta^2}) v_k = 0. \tag{2.5}$$

As an inflation we consider the power-law inflation $a(\eta) \approx (-\eta)^p$ $(= t^{p/(p+1)})$. The solution of Eq.(2.5) is then written as

$$f_k^I(\eta) = i\frac{\sqrt{\pi}}{2} e^{-ip\pi/2} (-\eta)^{1/2} H^{(1)}_{-p+1/2}(-k\eta), \tag{2.6}$$

where $H^{(1)}_{-p+1/2}$ is the Hankel function of the first kind of order $-p+\frac{1}{2}$. We assume that as a general initial condition, the mode function $v_k(\eta)$ is written as

$$v_k(\eta) = c_{g1} \; f_k^I(\eta) + c_{g2} \; f_k^{I*}(\eta), \tag{2.7}$$

where the coefficients $c_{g1}$ and $c_{g2}$ obey $|c_{g1}|^2 - |c_{g2}|^2 = 1$. Equation (2.7) describes a squeezed state. The important point is that the coefficients $c_{g1}$ and $c_{g2}$ do not change during inflation. In the conventional treatments, the Bunch-Davies state for the field $v_k(\eta)$, i.e. $c_{g1} = 1$ and $c_{g2} = 0$, because of the requisite condition that in the limit $\eta \to -\infty$, the field $v_k(\eta)$ must approach plane waves, e.g. $e^{-ik\eta}/\sqrt{2k}$ .

The power spectrum is defined as

$$< v_k(\eta)\, v_l^*(\eta) >= \frac{\pi m_P^2 a^2}{16 k^3} P_g \; \delta^3(\boldsymbol{k}\text{-}\boldsymbol{l}), \tag{2.8}$$

where $m_P$ is the Planck mass. The power spectrum of gravitational waves $P_g^{1/2}$ is then written as[4)]

$$P_g^{1/2} = \frac{4\sqrt{k^3}}{m_P\sqrt{\pi}\,a}|v_k|. \tag{2.9}$$

The series of the Hankel function $H^{(1)}_{-p+1/2}(z)$ at the limit of z$\to 0$ ($z$=-$k\eta$ ) is written as

$$H^{(1)}_{-p+1/2}(z)= z^{p-1/2}\left(-\frac{i2^{1/2-p}\sec p\pi}{\Gamma(1/2+p)}+\frac{i2^{-3/2-p}z^2\sec p\pi}{\Gamma(3/2+p)}-\frac{i2^{-9/2-p}z^4\sec p\pi}{\Gamma(5/2+p)}+o[z]^6\right)$$

$$+z^{-p-1/2}\left(\frac{2^{-1/2+p}(1+i\tan p\pi)z}{\Gamma(3/2-p)}-\frac{i2^{-5/2+p}(-i+\tan p\pi)z^3}{\Gamma(5/2-p)}-\frac{2^{-11/2+p}(1+i\tan p\pi)z^5}{\Gamma(7/2-p)}\right.$$

$$\left.+o[z]^7\right), \tag{2.10}$$

where $\Gamma(p+1/2)$ represents the Gamma function. In the present study, the term $z^{p-1/2}$ (=($-k\eta$)$^{p-1/2}$) is of the leading order. The power spectrum of the leading and next leading correction of |-$k\eta$| in the case of squeezed initial states can be written as

$$P_g^{1/2} = \left(\frac{2^{-p+1}}{\sqrt{\pi}}(-p)^p\frac{\Gamma(-p+1/2)}{\Gamma(3/2)}\frac{H}{m_P}\right)\Big|_{k=aH}\left(1-\frac{(-k\eta)^2}{2(1+2p)}\right)\times|c_{g1}\,e^{-ip\pi/2}+c_{g2}\,e^{ip\pi/2}|, \tag{2.11}$$

where $H$ is the Hubble expansion parameter, and $P_g^{1/2}$ is multiplied by a factor $\sqrt{2}$ for the two polarization states. We define the quantity $C_g(k)$ as

$$C_g(k) = c_{g1}\,e^{-ip\pi/2}+c_{g2}\,e^{ip\pi/2}. \tag{2.12}$$

This formula differs slightly from Hwang's formula[12)] due to the introduction of the term $e^{-ip\pi/2}$ into Eq. (2.6), as required in order that in the limit $\eta\to-\infty$, the field $v_k(\eta)$ must approach plane waves.[4)] The power spectrum of the gravitational waves is obtained by multiplying the familiar formation (i.e. $c_{g1}$ = 1 and $c_{g2}$ = 0) by the quantities $|C_g(k)|$.

## 2.2 Scalar perturbations

Next, the scalar perturbations are considered. The formula for the power spectrum of curvature perturbations for any initial conditions in inflation was obtained in the previous report.[9] The derivation is given here briefly in order clarify the difference between the tensor and scalar perturbations. Using the gauge-invariant potential *u*, the action for the scalar perturbations is written as[16]

$$S = \frac{1}{2}\int d\eta d^3x \{ (\frac{\partial u}{\partial \eta})^2 - c_s^2 \ (\nabla u)^2 + \frac{Z''}{Z} u^2 \ \}, \tag{2.13}$$

where $Z = \frac{a\dot{\phi}}{H}$, and $u = -Z\boldsymbol{R}$. The field $\phi$ is the inflaton field, $c_s$ is the sound velocity, *H* is the Hubble expansion parameter, and $\boldsymbol{R}$ is the curvature perturbation. Overdots represent derivatives with respect to *t*. The field $u(\eta, \boldsymbol{x})$ is expressed using annihilation and creation operators as follows.

$$u(\eta, x) = \frac{1}{(2\pi)^{3/2}}\int d^3k \{ \boldsymbol{a_k} \ u_{\mathrm{k}}(\eta) + \ a_{-k}{}^{\dagger} u^*{}_k(\eta)\} e^{-ikx} . \tag{2.14}$$

The field equation of $u_k$ is then derived as

$$\frac{d^2 u_k}{d\eta^2} + (c_s^2 \ k^2 - \frac{1}{Z}\frac{d^2 Z}{d\eta^2}) \ u_k \ = 0. \tag{2.15}$$

The solution $u_k$ satisfies the normalization condition $u_k\, du_k^* / d\eta\ -\ u_k^*\, du_k / d\eta\ = i$. Considering the power-law inflation $a(\eta) \approx (-\eta)^p\ (= t^{p/(p+1)})$, Eq. (2.15) can be rewritten as

$$\frac{d^2 u_k}{d\eta^2} + \left(k^2 - \frac{p(p-1)}{\eta^2}\right) u_k = 0, \tag{2.16}$$

where, in the scalar field case, $c_s^2 = 1$. Equation (2.6) is the solution of (2.16) because the equations for the scalar perturbations and tensor perturbations become the same in the case of the power-law inflation. Similar to the case for gravitational waves, as a general initial condition of the scalar perturbations, the mode function $u_k(\eta)$ is written as

$$u_k(\eta) = c_1\ f_k^I\ (\eta) + c_2\ f_k^{I*}(\eta), \tag{2.17}$$

where the coefficients $c_1$ and $c_2$ obey $|c_1|^2 - |c_2|^2 = 1$. Conventionally, the Bunch-Davies state is adopted for the field $u_k(\eta)$, i.e. $c_1 = 1$ and $c_2 = 0$.

The power spectrum of the curvature perturbations is defined as

$$< \boldsymbol{R}_k(\eta), \boldsymbol{R}_l^*(\eta) >= \frac{2\pi^2}{k^3} P_{\boldsymbol{R}} \delta^3(\boldsymbol{k}\text{-}\boldsymbol{l}), \tag{2.18}$$

where $\boldsymbol{R}_k(\eta)$ is the Fourier series of the curvature perturbation $\boldsymbol{R}$ ($u = -Z\,\boldsymbol{R}$). Then, the power spectrum $P_R{}^{1/2}$ is written as[4)]

$$P_R{}^{1/2} = \sqrt{\frac{k^3}{2\pi^2}}\, \left|\frac{u_k}{Z}\right|. \tag{2.19}$$

Using the approximation of the Hankel function (2.10), the power spectrum of the leading and next leading correction of $|-k\eta|$ in the case of squeezed initial states can be written as

$$P_{\mathcal{R}}^{1/2} = (2^{-p}(-p)^p \frac{\Gamma(-p+1/2)}{\Gamma(3/2)} \frac{1}{m_P^2} \frac{H^2}{|H'|})|_{k=aH} \; (1-\frac{(-k\eta)^2}{2(1+2p)}) \times |c_1 e^{-ip\pi/2} + c_2 e^{ip\pi/2}|$$

$$= (\frac{H^2}{2\pi\dot{\phi}})|_{k=aH} \; |c_1 e^{-ip\pi/2} + c_2 e^{ip\pi/2}|. \qquad (2.20)$$

The quantity $C(k)$ is defined as

$$C(k) = c_1 e^{-ip\pi/2} + c_2 e^{ip\pi/2}. \qquad (2.21)$$

Again, this formula differs slightly from that of Hwang[12)] for the same reasons as for gravitational case.

The power spectrum of the curvature perturbations is then obtained by multiplying the familiar formation (i.e. $c_1$ = 1 and $c_2$ = 0) by the quantity $|C(k)|$.

## 2.3 Scalar-tensor ratio and spectral index

The ratio of the power spectrum of the gravitational waves to that of the curvature perturbations, is obtained from Eqs. (2.11) and (2.20) as

$$R(k) \equiv P_g^{1/2} / P_{\mathcal{R}}^{1/2} = 4 \sqrt{\frac{p+1}{p}} \frac{|C_g(k)|}{|C(k)|}. \qquad (2.22)$$

From Eq. (2.22), if the coefficients $c_{g1}$ and $c_{g2}$ have the same values as $c_1$ and $c_2$, i.e., $|C_g(k)|/|C(k)|$ = 1, the ratio becomes the familiar one. In fact, this may appear to be the case for some models. The question then arises as to whether models for which $|C_g(k)|/|C(k)|$ is not equal to 1 exist. $R(k)$ is calculated later in this paper for a number of simplified cosmological models.

For the spectral indices of the tensor and the curvature perturbations, if we assume that $|C_g(k)|^2 \propto k^{\Delta n_g}$ and $|C(k)|^2 \propto k^{\Delta n}$, then $P_g$ and $P_{\mathcal{R}}$ can be written as $P_g \propto k^{n_g+\Delta n_g}$ and $P_{\mathcal{R}} \propto k^{n-1+\Delta n}$, where $n_g$ and $n$ are the ordinary contributions of the spectral indices (i.e. the case $c_{g1} = 1$ and $c_{g2} = 0$ or the case $c_1 = 1$ and $c_2 = 0$). If $\Delta n_g$ and $\Delta n$ are not near zero, then the appropriateness of the model becomes a serious problem. Later in this paper, $\Delta n_g$ and $\Delta n$, which represent the contribution of pre-inflation, are calculated for a range of simplified cosmological models.

## 3. Pre-inflationary cosmological models

A series of simplified models of pre-inflation will be used to demonstrate the proposed perturbations. Here, the pre-inflation model is considered to consist simply of a radiation-dominated period or a scalar-matter-dominated period. A simple cosmological model is assumed, as defined by

$$a^P = a_1(-\eta - \eta_j)^r,$$

$$a^I = a_2(-\eta)^p,$$

$$a^R = a_3(\eta - \eta_m),$$

$$a^M = a_4(\eta - \eta_n)^2, \tag{3.1}$$

where,

$$\eta_j = \left(\frac{r}{p} - 1\right)\eta_2,$$

$$\eta_m = (1 - \frac{1}{p})\eta_3 ,$$

$$\eta_n = -\eta_4 + 2(1 - \frac{1}{p})\eta_3 ,$$

$$a_1 = (\frac{p}{r})^r (-\eta_2)^{p-r} a_2 ,$$

$$a_3 = -p(-\eta_3)^{p-1} a_2 ,$$

$$a_4 = \frac{a_3}{2(\eta_4 - \eta_n)} , \tag{3.2}$$

The scale factor $a^I$ represents the ordinary inflation. If $p = -1$, the inflation is de-Sitter inflation, and if $p < -1$, the inflation is a power-law inflation. Inflation is assumed to begin at $\eta = \eta_2$ and end at $\eta = \eta_3$. The radiation-dominated period (in which the scale factor is $a^R$) follows, and the matter-dominated period (in which the scale factor is $a^M$) begins at $\eta = \eta_4$, leading to the current time $\eta_5$.

In pre-inflation, for the case $r = 1$, the scale factor $a^P$ indicates that the radiation–dominated period occurs, whereas for the case $r = 2$, the scale factor $a^P$ indicates that the scalar-matter-dominated period occurs. Here, the period of inflation is assumed to be sufficiently long, that is, in the plot of $|C(k)|$, $\eta_2$ is chosen as the time when perturbations of the current Hubble horizon size exceed the Hubble radius in inflation. For example, for $\eta_5 = 1.25\times10^{18}$, $\eta_4 = 1.66\times10^{16}$, and $\eta_3 = -6.63\times10^{-9}$, the value of $\eta_2$ depends on the value of $p$; if $p = -\frac{10}{9}$ ($a = t^{10}$), then $\eta_2 = -7.03\times10^{17}$, while if $p = -\frac{100}{99}$ ($a = t^{100}$), then $\eta_2 = -6.39\times10^{17}$.

4. Calculations of the power spectrum, tensor-to-scalar ratio and spectral index

In this section, the power spectra are calculated for a number of cosmological models in the tensor and scalar cases. Using the derived power spectra, the ratio of the gravitational waves to the curvature perturbations and spectral indices in the tensor and scalar cases are calculated, and the differences between two matching conditions in terms of these properties are investigated.

4.1 Radiation-dominated period before inflation

In the case of a radiation-dominated period before inflation, the scale factor becomes $a^P = a_1(-\eta - \eta_j)$, i.e., $r = 1$. A difference between the scalar and tensor perturbations (gravitational waves) occurs in the radiation-dominated period, i.e., the equations of the fields become different. In the case of gravitational waves, the solution of Eq. (2.5) is written as

$$f_k^R(\eta) = \frac{1}{\sqrt{2k}} e^{-ik(\eta+\eta_i)} . \tag{4.1}$$

On the other hand, in the case of curvature perturbations, the field equation $u_k$ can be written as Eq. (2.15) in the radiation-dominated period. In this case, $Z$ is written as $Z = a^{R1}\ ((\mathcal{H}^{\ 2} - (\mathcal{H}\,)')^{1/2}\ /\ (\sqrt{4\pi G}\ c_s\,\mathcal{H})$, where $\mathcal{H} = (a^{R1})'/a^{R1}$, and $G$ is Newton's gravitational constant.[11,16)] Fixing the value of $c_s^2$ at 1/3, the solution of Eq. (2.15) becomes

$$f_k^{SR}(\eta) = \frac{3^{1/4}}{\sqrt{2k}}\, e^{-ik(\eta+\eta_j)/\sqrt{3}} \,. \tag{4.2}$$

For simplicity, it is assumed that the mode functions of the radiation-dominated period can be written as Eq. (4.1) for the gravitational waves, and as Eq. (4.2) for the scalar perturbations. In the power-law inflation, Eqs. (2.5) and (2.15) become the same, and can be written as Eq. (2.6). Then, the general mode functions in inflation can be written as

$$v_k^I(\eta) = c_{g1}\ f_k^I(\eta) + c_{g2}\ f_k^{I*}(\eta), \tag{4.3}$$

$$u_k^I(\eta) = c_1\ f_k^I(\eta) + c_2\ f_k^{I*}(\eta). \tag{4.4}$$

In order to fix the coefficients $c_{g1}$, $c_{g2}$, $c_1$ and $c_2$, we use the matching condition that the mode function and first $\eta$-derivative of the mode function are continuous at a transition time $\eta = \eta_2$ ($\eta_2$ is the beginning of inflation). The coefficients $c_{g1}$, $c_{g2}$, $c_1$ and $c_2$ can then be calculated as

$$c_{g1} = -\frac{\sqrt{\pi}}{2\sqrt{2z}}\, e^{i(p\pi/2+2z/p)}\,((-1+p+iz)\,H^{(2)}_{-p+1/2}(z) + z\,H^{(2)}_{-p+3/2}(z)), \tag{4.5}$$

$$c_{g2} = -\frac{\sqrt{\pi}}{2\sqrt{2z}}\, e^{i(-p\pi/2+2z/p)}\,((-1+p+iz)\,H^{(1)}_{-p+1/2}(z) + z\ H^{(1)}_{-p+3/2}(z)), \tag{4.6}$$

$$c_1 = \frac{\sqrt{\pi}}{2 3^{3/4}\sqrt{2z}}\, e^{i(p\pi/2+2z/\sqrt{3}p)}\,((3\text{-}3p\text{-}i\sqrt{3}\,z)\,H^{(2)}_{-p+1/2}(z) - 3z\,H^{(2)}_{-p+3/2}(z)), \qquad (4.7)$$

$$c_2 = \frac{\sqrt{\pi}}{2 3^{3/4}\sqrt{2z}}\, e^{i(-p\pi/2+2z/\sqrt{3}p)}\,((3\text{-}3p\text{-}i\sqrt{3}\,z)\,H^{(1)}_{-p+1/2}(z) - 3z\;H^{(1)}_{-p+3/2}(z))$$

(4.8)

where $z = -k\eta_2$. The quantities $C_g(k)$ and $C(k)$ are then derived from Eqs. (2.12) and (2.21) as follows.

$$C_g(k) = -\frac{\sqrt{\pi}}{2\sqrt{2z}}\, e^{2iz/p}\,\{(-1+p+iz)(H^{(1)}_{-p+1/2}(z)+H^{(2)}_{-p+1/2}(z))+z(H^{(1)}_{-p+3/2}(z)+H^{(2)}_{-p+3/2}(z))\}$$

(4.9)

$C(k) =$

$$-\frac{\sqrt{\pi}}{2 3^{3/4}\sqrt{2z}}\, e^{2iz/\sqrt{3}p}\,\{(-3+3p+i\sqrt{3}\,z)(H^{(1)}_{-p+1/2}(z)+H^{(2)}_{-p+1/2}(z))+3z(H^{(1)}_{-p+3/2}(z)+H^{(2)}_{-p+3/2}(z))\}$$

(4.10)

The quantities $|C_g(k)|$ and $|C(k)|$ are plotted as a function of $z$ ($=-k\eta_2$) in Figs. 1 and 2 for the case $p = -10/9$. In the case of $z \to 0$ (super large scales), the quantities $C_g(k)$ and $C(k)$ are given by

$$C_g(k) \cong \frac{2^{-1+p}\sqrt{\pi}}{\Gamma(\frac{3}{2}-p)}\; e^{2iz/p}\; z^{-p}\; (1\text{-}p\text{-}iz), \qquad (4.11)$$

$$C(k) \cong \frac{2^{-1+p}\sqrt{\pi}}{3^{3/4}\Gamma(\frac{3}{2}-p)}\; e^{2iz/\sqrt{3}p}\; z^{-p}\; (3\text{-}3p\text{-}i\sqrt{3}\,z), \qquad (4.12)$$

In the case of $z \to 0$, both $C_g(k)$ and $C(k)$ become zero. Next, we consider the case $z \to \infty$ (from large scales to small scales). The quantities $C_g(k)$ and $C(k)$ are then approximately

$$|C_g(k)| \cong (1+\frac{p^2(p-1)^2}{4z^2})^{1/2}, \tag{4.13}$$

$$|C(k)| \cong \{\frac{1}{\sqrt{3}}(2+\cos(p\pi+2z))\}^{1/2}, \tag{4.14}$$

The behavior differs between the tensor and scalar cases, i.e., $C_g(k)$ becomes 1 and $C(k)$ oscillates around $\sqrt{2}/3^{1/4} \cong 1.07$.

From $C_g(k)$ and $C(k)$, the ratio of gravitational waves to the curvature perturbations on the power spectrum $R(k) = 4\sqrt{(p+1)/p}\ R_c(k)$ can be calculated. In the case of $z \to 0$ (super large scales), $R_c(k)$ is derived as

$$R_c(k) = \frac{|C_g(k)|}{|C(k)|} \cong \sqrt{\frac{\sqrt{3}(1-2p+p^2+z^2)}{3-6p+3p^2+z^2}}, \tag{4.15}$$

which corresponds to $3^{-1/4}$ ($\cong 0.76$) when $p$ = –10/9. In the case of |z|>>1, $R_c(k)$ becomes

$$R_c(k) \cong (\frac{\sqrt{3}}{2+Cos(p\pi+2z)})^{1/2}, \tag{4.16}$$

This ratio $R_c(k)$ oscillates around $\sqrt{2}/3^{1/4} \cong 1.07$, numerically $0.760 \leq R_c(k) \leq 1.316$, and is largely independent of $p$. The ratio $R_c(k)$ is plotted as a function of $z$ in Fig. 3 for the case $p$ = –10/9 to examine the overall behavior of the ratio.

Finally, we consider the spectral indices of the tensor and curvature perturbations. Here, only the difference from the normal values is considered, i.e., $\Delta n_g$ and $\Delta n$. The spectral indices for the case of scalar and tensor are obtained using Eqs. (4.9) and (4.10). In the case $z \to 0$, $\Delta n_g$ and $\Delta n$ become -2*p*, which is non-zero. However, if inflation is sufficiently long, this behavior becomes less important. In the case |*z*|>1 (from large scales to small scales), the values of $\Delta n_g$ and $\Delta n$ are zero.

4.2 Scalar-matter-dominated period before inflation

When the period before inflation is scalar-matter-dominated, in which the scalar-matter is the inflaton field $\phi$, the scale factor becomes $a^{R1} = a_1(-\eta - \eta_j)^2$, i.e., $r = 2$. In this case, Eqs. (2.5) and (2.15) become the same, and consequently the quantities $C_g(k)$ and $C(k)$ are the same form. Using a similar procedure as in section 4.1, the quantities $C_g(k)$ and $C(k)$ become

$$C_g(k) = C(k) = \frac{-i\sqrt{\pi}}{8\sqrt{2z^3}} e^{2iz/p} \{(p^2 + 4z(i+z) - 2p(1+iz))(H^{(1)}_{-p+1/2}(z) + H^{(2)}_{-p+1/2}(z))$$

$$+ 2(p-2iz)z(H^{(1)}_{-p+3/2}(z) + H^{(2)}_{-p+3/2}(z))\}. \tag{4.17}$$

In order to compare the case of the radiation-dominated period with the case of the scalar-matter-dominated period, |*C*(*k*)| as given above is plotted as a function of *z* in Fig. 4 for the case $p = -10/9$. As the quantities $C_g(k)$ and $C(k)$ are the same, the ratio of scalar to tensor perturbations *R*(*k*) is the ordinary value, i.e., $R_c(k) = 1$. The spectral indices $\Delta n_g$

and $\Delta n$ are equal in this case. For $z \to 0$, $\Delta n_g$ $(=\Delta n)$ becomes a constant (=–2($p$+1)) from Eq. (4.17), whereas for |z|>>1, $\Delta n_g$ $(=\Delta n)$ oscillates and approaches zero. The details are shown in Table 1.

The behavior of |$C$($k$)| for |z|>>1 differs between the radiation-dominated and matter-dominated assumptions; |$C$($k$)| oscillates assuming a radiation-dominated pre-inflation period, |$C$($k$)| $\cong$ 1 for a matter-dominated pre-inflation period. Different behavior is also observed for the spectral indices in the case of $z \to 0$.

4.3. Matching condition

Here, the matching condition for the scalar perturbations is considered. One of two matching conditions is used in sections 4.1 and 4.2. One is a matching condition in which the gauge potential and its first $\eta$ -derivative are continuous at the transition point. This matching condition allows the initial condition of pre-inflation to be decided rationally, i.e., in the limit $\eta \to -\infty$, the field $u_k(\eta)$ approaches plane waves. The other is Deruelle and Mukhanov's [13)] matching condition for cosmological perturbations, which requires that the transition occurs on a hyper-surface of constant energy.

Using Deruelle's matching condition that $\Phi$ and $\zeta$ (***R***) are continuous at the transition time $\eta_2$, the quantity |$C$($k$)|, the ratio of the power spectrum of gravitational waves to that of the curvature perturbations, and the spectral indices of the curvature

perturbations are calculated, and the difference between these two matching conditions are investigated. The parameters $\Phi$ and $\zeta$ can be written as follows. [11,16)]

$$\Phi = (-\boldsymbol{H}^2 + \boldsymbol{H}') \left(\frac{u}{Z}\right)' / (c_s^2 k^2 \boldsymbol{H}), \tag{4.18}$$

$$\zeta = (\boldsymbol{H}\Phi' - \Phi \boldsymbol{H}' + 2\boldsymbol{H}^2 \Phi)/(\boldsymbol{H}^2 - \boldsymbol{H}'), \tag{4.19}$$

where $Z$ is written as $a(\boldsymbol{H}\ \boldsymbol{H}^2 - \boldsymbol{H}')^{1/2}/(c_s\sqrt{4\pi G}\ \boldsymbol{H})$. As $\Phi$ and $\zeta$ can be written in terms of $u_k(\eta)$, the coefficients $c_1$ and $c_2$ are obtained as follows: In the period of pre-inflation, the mode function $u_k(\eta)$ is derived from Eq. (2.15), and $\Phi^R$ and $\zeta^R$ can be obtained using the relations (4.18) and (4.19). On the other hand, in inflation, $u_k(\eta)$ is expressed as $c_1 f_k^I(\eta) + c_2 f_k^{I*}(\eta)$, and this is used to calculate $\Phi^I$ and $\zeta^I$. From the relations $\Phi^R(\eta_2) = \Phi^I(\eta_2)$ and $\zeta^R(\eta_2) = \zeta^I(\eta_2)$, the coefficients $c_1$ and $c_2$ can be fixed.

Here, the matching condition that $\Phi$ and $\zeta$ (***R***) are continuous at the transition point is adopted. However, the matching condition of Deruelle and Mukhanov must be written such that $\Phi$ and $\zeta + k^2\Phi/(3(\boldsymbol{H}^2 - \boldsymbol{H}'))$ are continuous at the transition point. In the present case ($k\eta_2 = -z$), the value of $\zeta$ becomes smaller than that of the $k^2$ term, and the $k^2$ term dominates at $|z|>>1$. However, the coefficients $c_1$ and $c_2$ cannot be fixed using $\Phi$ and the $k^2$ term. Thus, the matching condition that $\Phi$ and $\zeta$ (***R***) are continuous at the transition point is adopted here. As the calculation of the coefficients $c_1$ and $c_2$ is similar to that in previous sections, only the results are given here.

In the case of the radiation-dominated period before inflation, $|C(k)|$ can be written from Eq. (2.21) as

$$C(k) = \frac{\sqrt{\pi}}{4 3^{3/4} \sqrt{p(p+1)}\sqrt{z}} e^{iz/\sqrt{3}p} \{(\sqrt{3} + 4\sqrt{3}\, p^2 - p(\sqrt{3} + 6iz))$$

$$\times (H^{(1)}_{-p+1/2}(z) + H^{(2)}_{-p+1/2}(z)) + \sqrt{3}(1+p)z(H^{(1)}_{-p+3/2}(z) + H^{(2)}_{-p+3/2}(z))\}. \quad (4.20)$$

For $z \to 0$, $|C(k)|$ becomes zero, while for $|z|>>1$, $|C(k)|$ oscillates around $\sqrt{-p-1}/(\sqrt{-2p}\ 3^{1/4}) \le |C(k)| \le \sqrt{-2p}\ 3^{1/4}/\sqrt{-p-1}$, which corresponds to a numerical range of $0.170 \le |C(k)| \le 5.89$ when $p = -10/9$. The variation in $|C(k)|$ as a function of $z$ is shown in Fig. 5 for $p = -10/9$. Comparing Fig. 2 with Fig. 5, some differences between the matching conditions are apparent.

Calculating the tensor-to-scalar ratio $R_c(k)$ from $C_g(k)$ and $C(k)$, for $z \to 0$ (super large scales) $R_c(k)$ becomes 0.196 when $p = -10/9$, while for $|z|>>1$, $R_c(k)$ oscillates in the range $0.170 \le R_c(k) \le 5.89$. The ratio $R_c(k)$ is plotted as a function of $z$ in Fig. 6 for $p = -10/9$ over the range $0 \le z \le 10$. The deviation of the spectral index $\Delta n$ becomes $-2p$ for $z \to 0$, and zero for $|z|>>1$.

From a comparison of the two matching conditions, the behavior of $C(k)$ and $R_c(k)$ is similar but the values are different, whereas $\Delta n$ has the same value for both matching conditions. The details are listed in Table 1.

For the case of the scalar-matter-dominated period before inflation, $C(k)$ is written from Eq. (2.21) as

$$C(k) = \frac{\sqrt{\pi}}{16\sqrt{3p(p+1)}\sqrt{z^3}}\, i\, e^{2iz/p}$$

$$\{(p^3 + p^2(-4-2iz) - 8iz + p(4+8iz-12z^2))(H^{(1)}_{-p+1/2}(z) + H^{(2)}_{-p+1/2}(z))$$

$$-4(1+p)(p-2iz)z(H^{(1)}_{-p+3/2}(z) + H^{(2)}_{-p+3/2}(z))\}. \qquad (4.21)$$

In the case of $z \to 0$, $|C(k)|$ becomes zero, and in the case of $|z|>>1$, $|C(k)|$ oscillates around $\sqrt{-2(p+1)}/(\sqrt{-3p}) \le |C(k)| \le \sqrt{-3p}\, 3^{1/4}/\sqrt{-2(p+1)}$, corresponding to a numerical range of $0.258 \le |C(k)| \le 3.873$ when $p = -10/9$. As the behavior of $|C(k)|$ differs between the two matching conditions for $|z|>>1$, $|C(k)|$ is plotted as a function of $z$ $(=-k\eta_2)$ in Fig. 7 for $p = -10/9$. Again, in comparison with Fig. 3, some differences between the matching conditions are apparent. The tensor-to-scalar ratio $R_c(k)$ for $z \to 0$ (super large scales) is 0.277 when $p = -10/9$, and at $|z|>>1$ oscillates within the numerical range $0.258 \le R_c(k) \le 3.873$. Thus, the ratio differs between the two matching conditions, similar to the case for the radiation-dominated pre-inflationary period. The deviation of the spectral index $\Delta n$ for $z \to 0$ is $-2(p+1)$, while for $|z|>>1$, $\Delta n$ is zero.

Comparing the two matching conditions in the case of the scalar-matter-dominated pre-inflation period, the behavior of $|C(k)|$ for the two cases differs when $|z|>>1$, i.e., oscillation or a constant value of 1. Although the ratio differs, the behavior of the spectral indices is the same.

The results for these 4 cases (two models and two matching conditions) are summarized in Table 1.

## 5. Present power spectrum of the gravitational waves

The present power spectrum of the gravitational waves is calculated here using the simple cosmological model outlined in section 3. In inflation, which starts at $\eta_2$, the mode function of the gravitational waves is written as Eq. (2.7), where the coefficients $c_{g1}$ and $c_{g2}$ represent the effect of pre-inflation. The radiation-dominated period follows, starting at $\eta_3$. The mode function of the gravitational waves in this period is written as

$$v_k^R(\eta) = c_{g3}\, f_k^R(\eta - \eta_m) + c_{g4}\, f_k^{R*}(\eta - \eta_m). \tag{5.1}$$

In this case, the mode function $f_k^R(\eta)$ is the similar to Eq. (4.1). Finally, in the matter-dominated period, which starts at $\eta_4$, the mode function of the gravitational waves is written as

$$v_k^M(\eta) = c_{g5}\; f_k^M(\eta) + c_{g6}\; f_k^{M*}(\eta), \tag{5.2}$$

where

$$f_k^M(\eta) = \frac{1}{\sqrt{2k}}\left(1 - \frac{i}{k(\eta - \eta_n)}\right) e^{-ik(\eta - \eta_n)}. \tag{5.3}$$

Using the matching condition at $\eta_3$ and $\eta_4$, the coefficients $c_{g5}$ and $c_{g6}$ are derived as

$$c_{g5} = c_1\{\alpha_1\; f_k^I(\eta_3) + \alpha_2\; f_k^{I\prime}(\eta_3)\} + c_2\{\alpha_1\; f_k^{I*}(\eta_3) + \alpha_2\; f_k^{I*\prime}(\eta_3)\},$$

$$c_{g6} = c_1\{\beta_1\; f_k^I(\eta_3) + \beta_2\; f_k^{I\prime}(\eta_3)\} + c_2\{\beta_1\; f_k^{I*}(\eta_3) + \beta_2\; f_k^{I*\prime}(\eta_3)\}, \tag{5.4}$$

where

$$\alpha_1 = (-f_k^{R*\prime}(\eta_3) f_k^{R\prime}(\eta_4) + f_k^{R\prime}(\eta_3) f_k^{R*\prime}(\eta_4)) f_k^{M*}(\eta_4)$$

$$+ (f_k^{R*\prime}(\eta_3) f_k^{R}(\eta_4) - f_k^{R\prime}(\eta_3) f_k^{R*}(\eta_4)) f_k^{M*\prime}(\eta_4),$$

$$\alpha_2 = (f_k^{R*}(\eta_3) f_k^{R\prime}(\eta_4) - f_k^{R}(\eta_3) f_k^{R*\prime}(\eta_4)) f_k^{M*}(\eta_4)$$

$$+ (-f_k^{R*}(\eta_3) f_k^{R}(\eta_4) + f_k^{R}(\eta_3) f_k^{R*}(\eta_4)) f_k^{M*\prime}(\eta_4),$$

$$\beta_1 = -\{(-f_k^{R*\prime}(\eta_3) f_k^{R\prime}(\eta_4) + f_k^{R\prime}(\eta_3) f_k^{R*\prime}(\eta_4)) f_k^{M}(\eta_4)$$

$$+ (f_k^{R*\prime}(\eta_3) f_k^{R}(\eta_4) - f_k^{R\prime}(\eta_3) f_k^{R*}(\eta_4)) f_k^{M\prime}(\eta_4)\},$$

$$\beta_2 = -\{(f_k^{R*}(\eta_3) f_k^{R\prime}(\eta_4) - f_k^{R}(\eta_3) f_k^{R*\prime}(\eta_4)) f_k^{M}(\eta_4)$$

$$+ (-f_k^{R*}(\eta_3) f_k^{R}(\eta_4) + f_k^{R}(\eta_3) f_k^{R*}(\eta_4)) f_k^{M\prime}(\eta_4)\}. \quad (5.5)$$

As $|z_3| << 1$ ($z_3 = k\eta_3$), using the approximation (2.10), we obtain the following relations.

$$f_k^{I}(\eta_3) \cong e^{-ip\pi/2} |\hat{f}_k^{I}(\eta_3)|, \quad f_k^{I*}(\eta_3) \cong e^{ip\pi/2} |\hat{f}_k^{I}(\eta_3)|,$$

$$f_k^{I\prime}(\eta_3) \cong e^{-ip\pi/2} |\hat{f}_k^{I\prime}(\eta_3)|, \quad f_k^{I*\prime}(\eta_3) \cong e^{ip\pi/2} |\hat{f}_k^{I\prime}(\eta_3)|. \quad (5.6)$$

Using these relations, the coefficients $c_{g5}$ and $c_{g6}$ are obtained as

$$c_{g5} \cong (c_{g1} e^{-ip\pi/2} + c_{g2} e^{ip\pi/2})(\alpha_1 |\hat{f}_k^{I}(\eta_3)| + \alpha_2 |\hat{f}_k^{I\prime}(\eta_3)|),$$

$$c_{g6} \cong (c_{g1} e^{-ip\pi/2} + c_{g2} e^{ip\pi/2})(\beta_1 |\hat{f}_k^{I}(\eta_3)| + \beta_2 |\hat{f}_k^{I\prime}(\eta_3)|). \quad (5.7)$$

Finally, the mode function $v_k^M(\eta)$ can be written as

$$v_k^M(\eta) \cong (c_{g1} e^{-ip\pi/2} + c_{g2} e^{ip\pi/2})$$

$$\times \{(\alpha_1 |\hat{f}_k^{I}(\eta_3)| + \alpha_2 |\hat{f}_k^{I\prime}(\eta_3)|) f_k^{M}(\eta) + (\beta_1 |\hat{f}_k^{I}(\eta_3)| + \beta_2 |\hat{f}_k^{I\prime}(\eta_3)|) f_k^{M*}(\eta)\}.$$

(5.8)

The present power spectrum $P_g^{1/2}(\eta_5)$ is then written as

$$P_g^{1/2}(\eta_5) = \frac{4\sqrt{k^3}}{m_P\sqrt{\pi}\,a(\eta_5)}|v_k(\eta_5)|.$$

$$\cong (P_g^{normal}(\eta_5))^{1/2}|c_{g1}e^{-ip\pi/2} + c_{g2}e^{ip\pi/2}|, \qquad (5.9)$$

where $(P_g^{normal}(\eta_5))^{1/2}$ indicates the case of $c_{g1} = 1$ and $c_{g2} = 0$ (normal case).

This result is very important: the present power spectrum of the gravitational waves can be obtained by multiplying the familiar formation by the factor $|C_g(k)|$ with an approximation. Therefore, the contribution of pre-inflation on the power spectrum applies only to the quantities $|C_g(k)|$.

5.1 Radiation-dominated period before inflation

The present power spectrum of gravitational waves in the case of the radiation-dominated pre-inflation period is calculated here using the simple cosmological model (3.1). With $C_g(k)$ in the leading term of $z_3$, the power spectrum can be calculated as

$$P_g^{1/2}(\eta_5) \cong \frac{4k^{3/2}}{\sqrt{\pi}\,m_P}$$

$$|- \frac{(-z_3)^{-p+1}}{\sqrt{k\pi}\,z_4^2(z\ \ + z_4)}\ C_g(k)\,(i\,2^{-9/2-p}p(e^{2iz_4}(-i + z\ \ + z_4)(-1 + e^{2iz_4} + 4i\,z_4 + 8\,z_4^2)$$

$$- e^{2iz_5}(i + z\ \ + z_4)(1 + e^{2iz_4}(-1 - 4i\,z_4 + 8\,z_4^2))\,\Gamma(1/2 - p)| \qquad (5.10)$$

where $z = k\eta_5$, $z_4 = k\eta_4$, and $z_3 = -k\eta_3$. In this case, a($\eta_5$) = 1 is used. Many papers treat this power spectrum for the case of $C_g(k)$ = 1 and de-Sitter inflation. For the case of a radiation-dominated period before inflation, the quantity $C_g(k)$ is as shown in Eq. (4.9). The spectrum is plotted as a function of $z(=k\eta_5)$ in Fig. 8 for two cases; the normal case ($c_1$= 1 and $c_2$ = 0) (*Case 1*), and in the case of inflation before the radiation-dominated period (*Case 2*). When $z$<1, the present power spectrum for *Case 1* is constant, but becomes small for *Case 2*. When $z$>1, the two cases are not appreciably different.

## 6. Summary

The influence of the initial condition in inflation on the scalar and tensor perturbations was investigated. Assuming a squeezed initial state, the power spectra of the gravitational waves and curvature perturbations were calculated. The derived formulae represent the commonly used formulae multiplied by the factor $|C_g(k)|$ or $|C(k)|$. The effect of the squeezed initial state in inflation on the tensor-to-scalar ratio and spectral index were also investigated using the derived formulae. Assuming some simplified pre-inflationary cosmological models, the power spectra of the gravitational waves and curvature perturbations were calculated and used to obtain the ratio and spectral index. These properties were examined for two matching conditions. The results, as tabulated in Table 1, reveal that some differences exist between the cosmological models and matching conditions, and although not large, cannot be ignored. Here, when we plot the quantities, the value (p=-10/9) is adopted, but for other values (for example, p=-100/99) similar results were derived.

The present power spectrum of gravitational waves for any initial conditions was derived. The proposed formula is quite simple, being the familiar formation multiplied by a factor indicating the contribution of the initial condition. For a simplified model in which a radiation-dominated period exists before inflation, this power spectrum differs from that determined based on the Bunch-Davies vacuum as the initial condition of inflation at super large scales. In the ordinary case, the present power spectrum of the gravitational waves becomes a constant, but for a radiation-dominated period before inflation becomes zero. This difference is eliminated if inflation is sufficiently long.

Recent observations suggest that matter associated with a non-zero cosmological constant (dark energy) dominate in the present day. The effect of a non-zero cosmological constant on the power spectrum of background gravitational waves was investigated in Ref. 3, in which it was indicated that the power spectrum of background gravitational waves under a non-zero cosmological constant is 0.3–0.6 times that of models assuming a cosmological constant of zero on large scales. As this estimate can be applied to our case, the present power spectrum of the gravitational waves may become smaller.

The physics of pre-inflation has been investigated intensively in recent years, and many pre-inflation models have been proposed. The formulae derived here represent a means of determining whether a pre-inflation model is appropriate. In the future, the present formulation will be applied to calculation of the power spectra of the scalar and tensor perturbations for more realistic models.

Acknowledgments

The author would like thank the staff of Osaka Electro-Communication Junior College for their invaluable discussions.

*) On leave from Osaka Electro-Communication Junior College

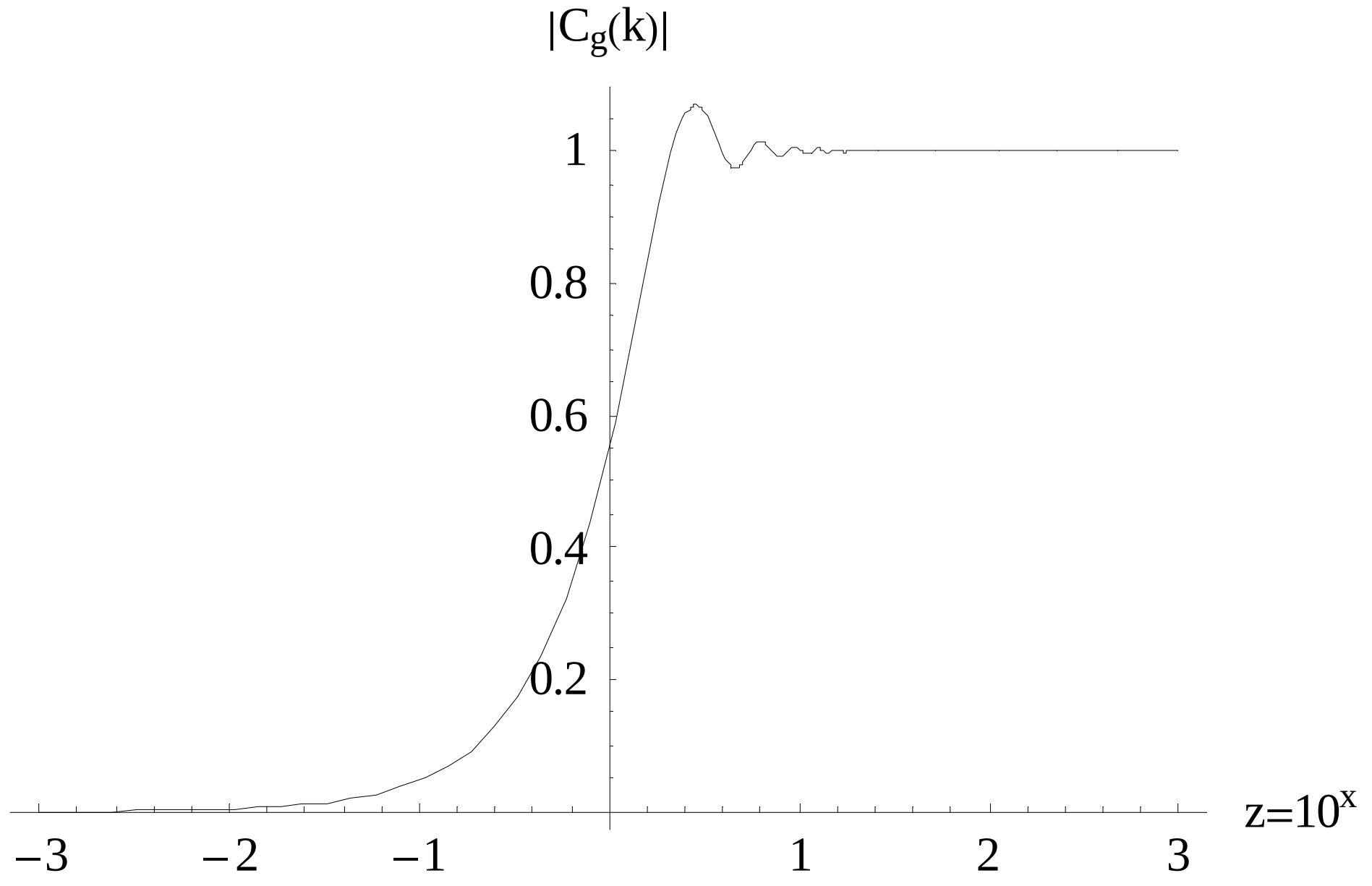

Fig. 1. Factor $|C_g(k)|$ as a function of $z$ ($=-k\eta_2$) for $10^{-3} \le z \le 10^3$ for tensor perturbations in the case of a radiation-dominated period before inflation ($p = -10/9$)

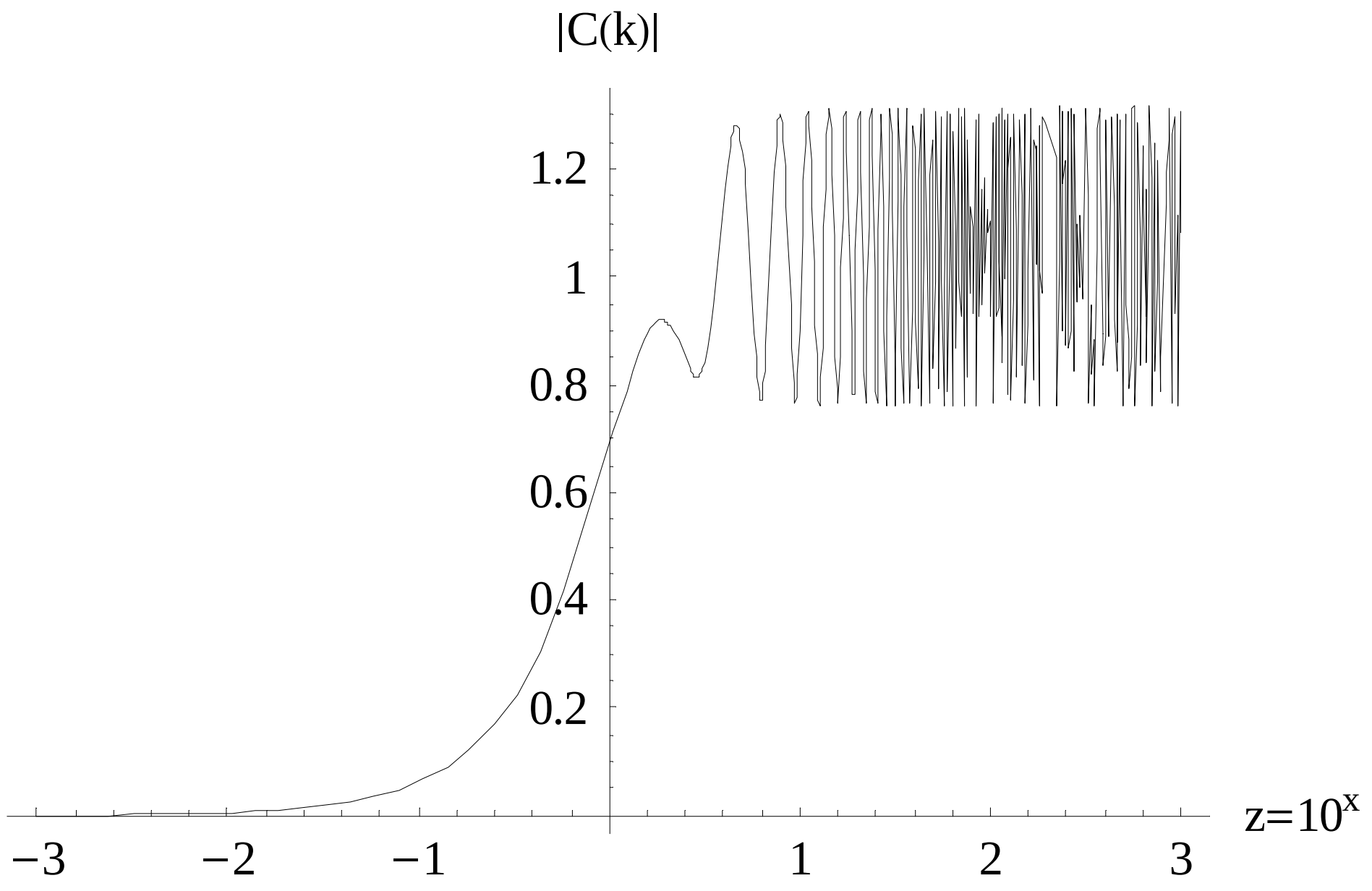

Fig. 2. Factor $|C(k)|$ as a function of $z$ $(=-k\eta_2)$ for $10^{-3} \le z \le 10^{3}$ for scalar perturbations in the case of radiation-dominated period before inflation ($p$ = –10/9)

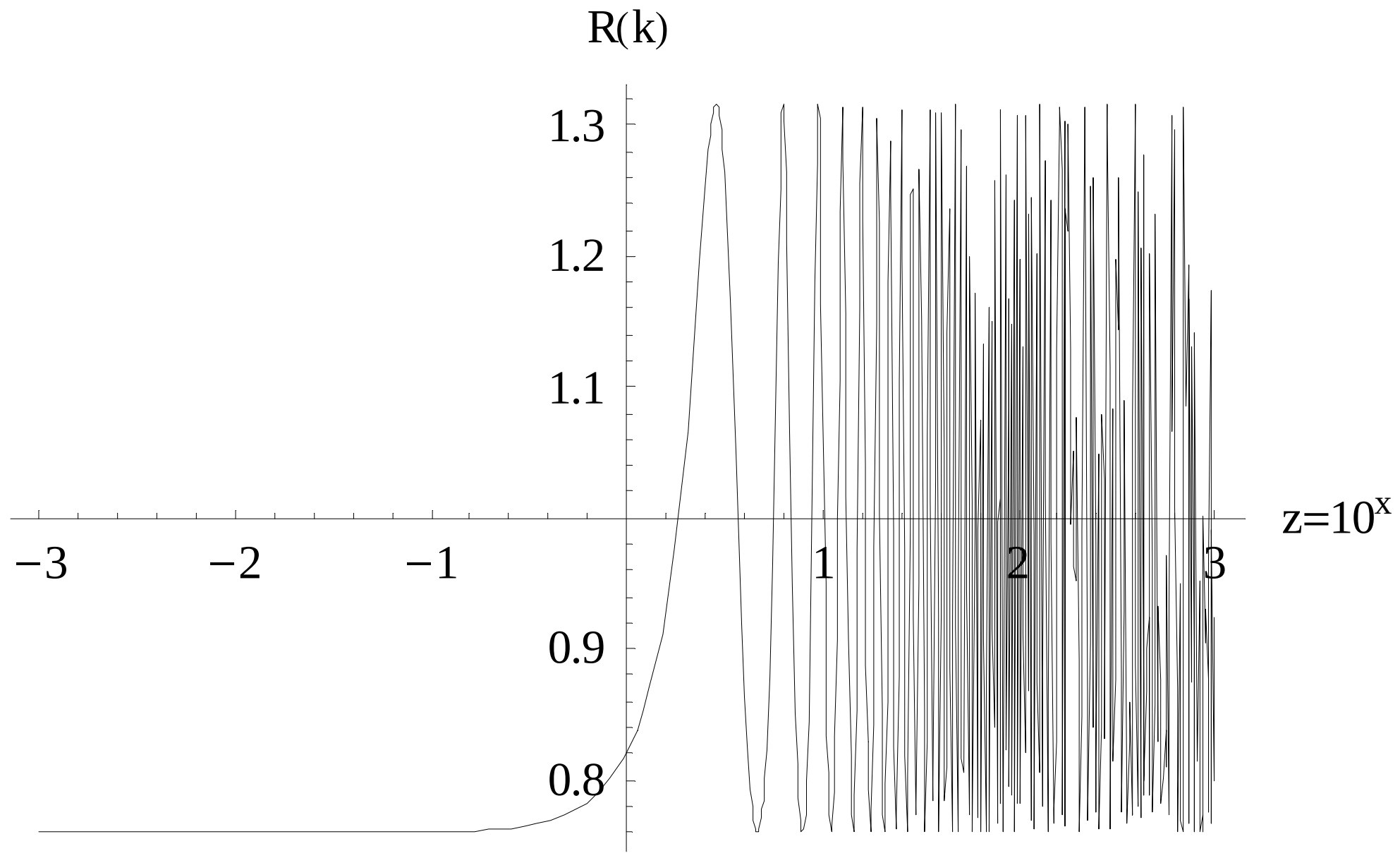

Fig. 3. Ratio $R_c = |C_g(k)|/|C(k)|$ in the case of radiation-dominated period before inflation ($p$ = –10/9)

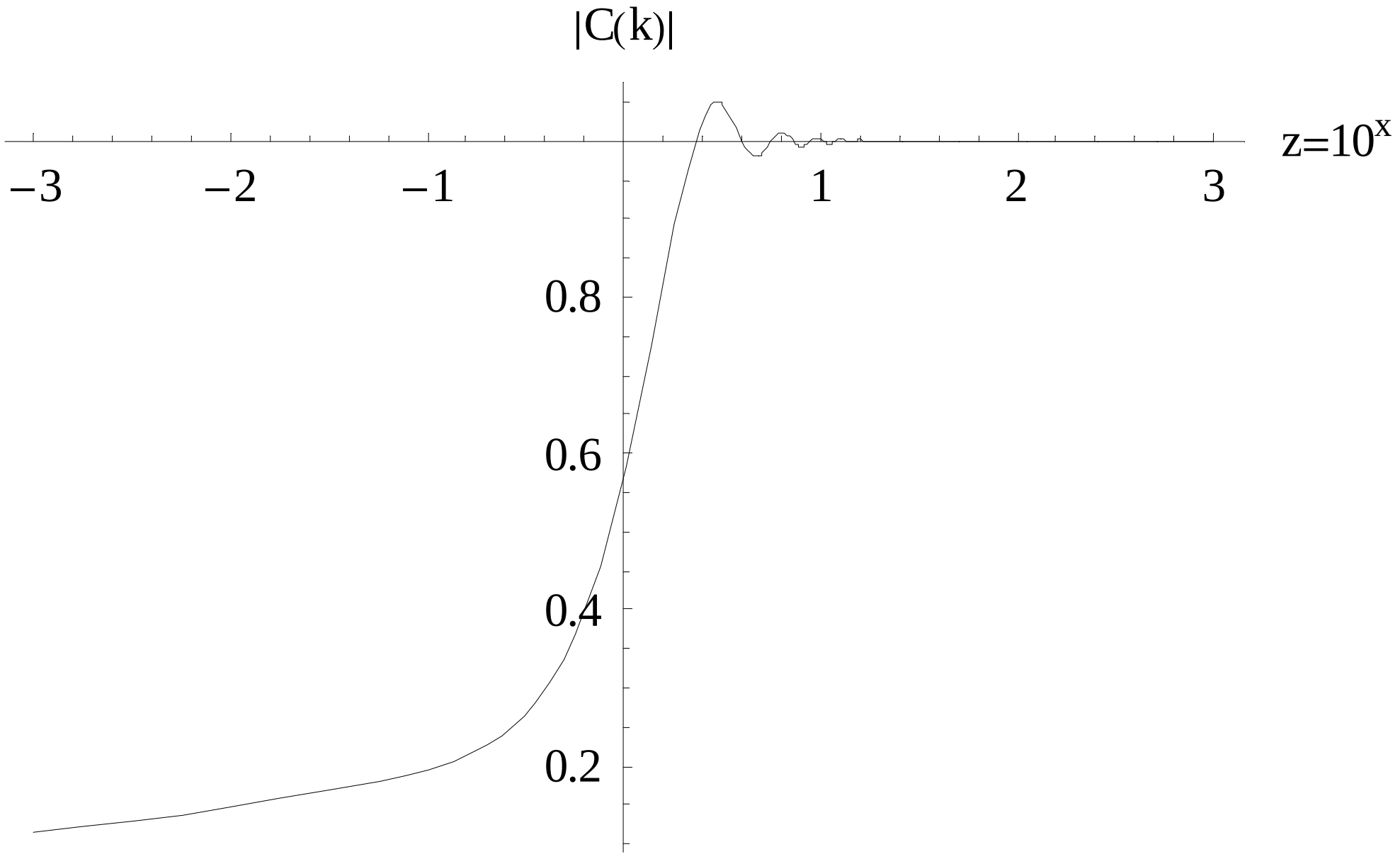

Fig. 4. Factor $|C(k)|$ $(=|C_g(k)|)$ in the case of scalar-matter-dominated period before inflation ($p$ = –10/9)

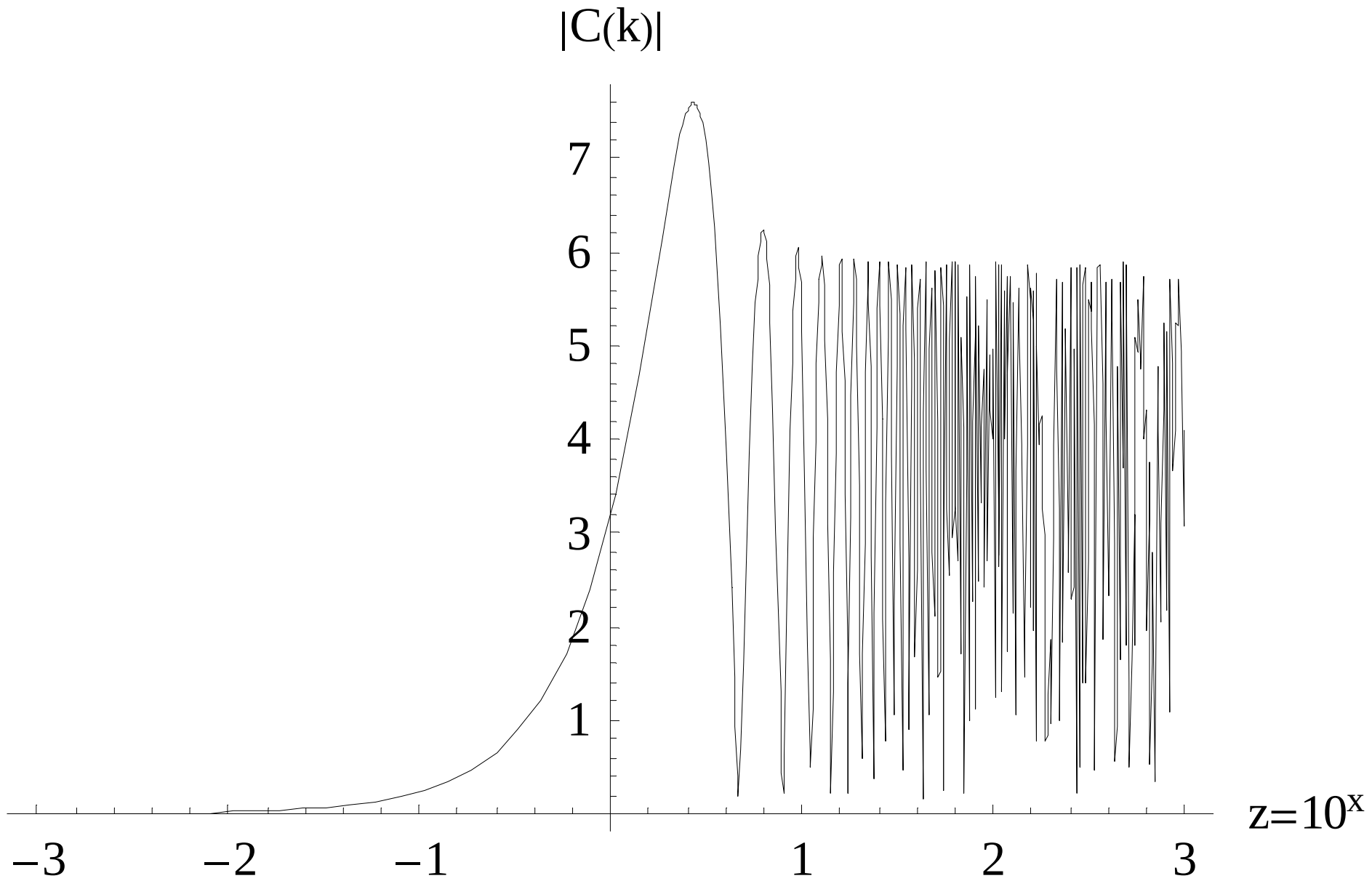

Fig. 5. Factor $|C(k)|$ using Deruell's matching condition in the case of radiation-dominated period before inflation ($p$ = –10/9).

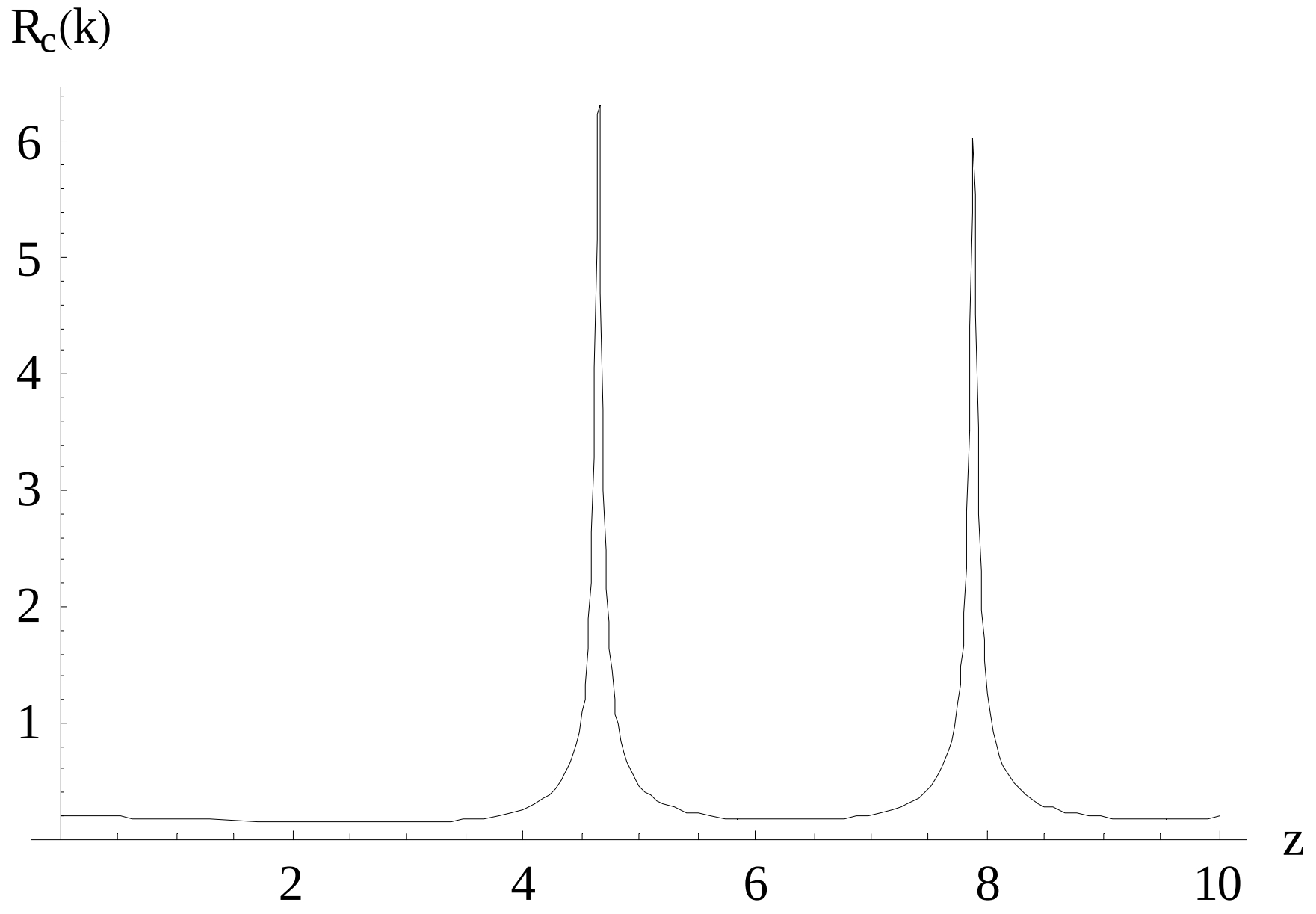

Fig. 6. Ratio $R_c = |C_g(k)|/|C(k)|$ using Deruell's matching condition in the case of radiation-dominated period before inflation ($1 \leq z \leq 10$ , $p$ = –10/9)

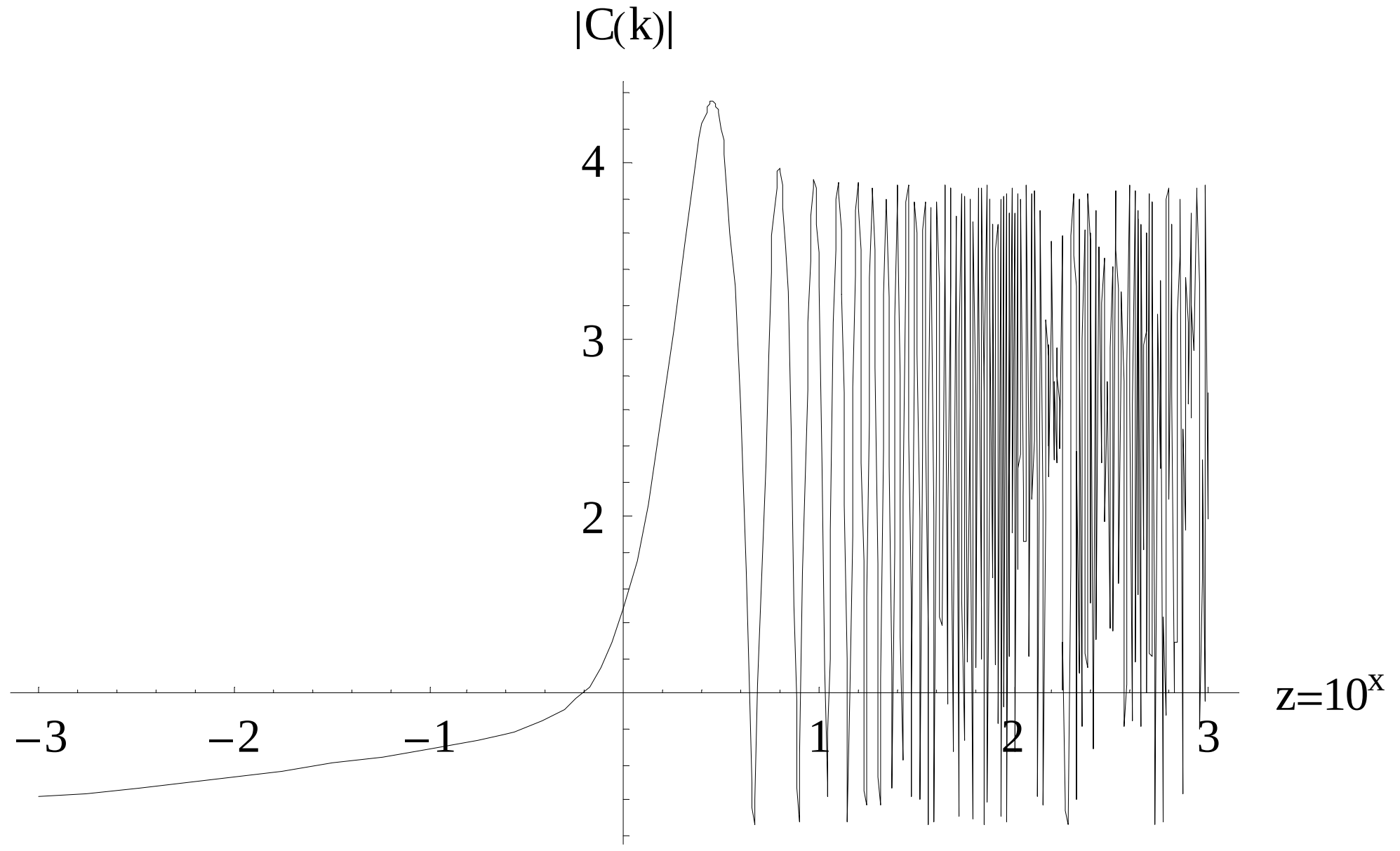

Fig. 7. Factor $|C(k)|$ using Deruell's matching condition in the case of scalar-matter-dominated period before inflation ($p = -10/9$).

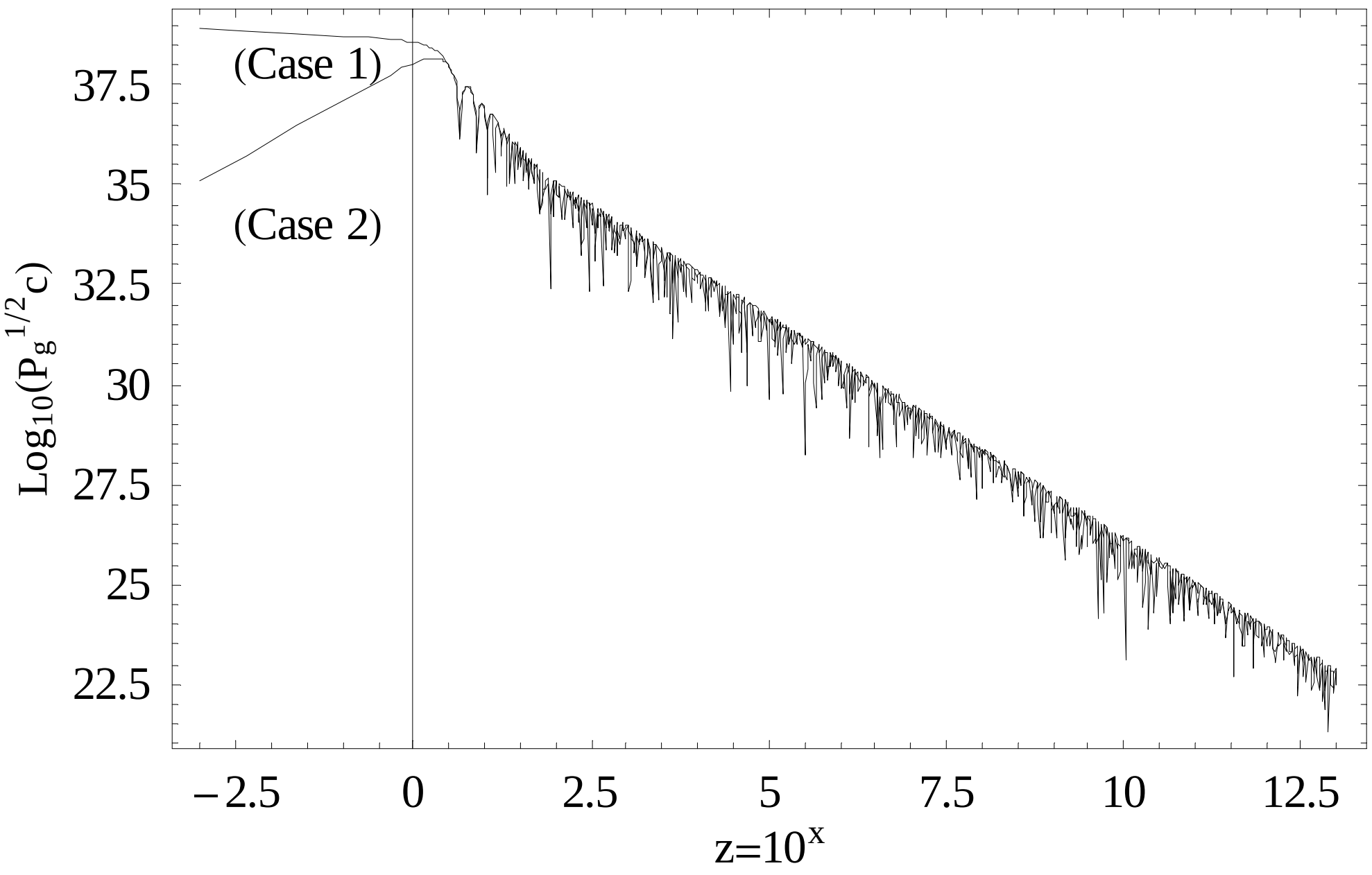

Fig. 8. Power spectrum of gravitational waves as a function $z$ (= $k\eta_5$) in the ordinary case (*Case 1*) and in the case of inflation before the radiation-dominated period (*Case 2*) in the range of $10^{-2} \le z \le 10^{13}$, where $c = \sqrt{\pi} M_P/4$.

Case of Radiation

| Matching condition | | $u_k$ and $u_k'$ | $\Phi$ and $\zeta$ |
|---|---|---|---|
| $\lvert C(k)\rvert$ | $z \to 0$ | $\dfrac{2^{-1+p}\sqrt{\pi}\,z^{-p}(3-3p)}{3^{3/4}\Gamma(\frac{3}{2}-p)}$ | $\dfrac{2^{-3/2+p}\sqrt{\pi}\,z^{-p}(4p^2-p+1)}{3^{1/4}\sqrt{p(p+1)}\Gamma(\frac{3}{2}-p)}$ |
| | $\lvert z\rvert >> 1$ | $\left(\dfrac{2+Cos(p\pi+2z)}{\sqrt{3}}\right)^{1/2}$ | $\sqrt{\dfrac{1+2p+13p^2+(1+2p-11p^2)Cos(p\pi+2z)}{4\sqrt{3}p(1+p)}}$ |
| Range ($p$=-10/9) | | $0.760 \le \lvert C(k)\rvert \le 1.316$ | $0.170 \le \lvert C(k)\rvert \le 5.89$ |
| $R_c(k)$ | $z \to 0$ | $3^{-1/4} \cong 0.76$ | $\dfrac{\sqrt{2}\,3^{1/4}(1-p)\sqrt{p(p+1)}}{4p^2-p+1} \cong 0.196$ |
| | $\lvert z\rvert >> 1$ | $\left(\dfrac{\sqrt{3}}{2+Cos(p\pi+2z)}\right)^{1/2}$ | $\sqrt{\dfrac{4\sqrt{3}p(1+p)}{1+2p+13p^2+(1+2p-11p^2)Cos(p\pi+2z)}}$ |
| Range ($p$=-10/9) | | $0.760 \le R_c(k) \le 1.316$ | $0.170 \le R_c(k) \le 5.89$ |
| $\Delta n$ | $z \to 0$ | $-2p$ | $-2p$ |
| | $\lvert z\rvert >> 1$ | 0 | 0 |

Case of Scalar-Matter

| Matching condition | | $u_k$ and $u_k'$ | $\Phi$ and $\zeta$ |
|---|---|---|---|
| $\lvert C(k)\rvert$ | $z \to 0$ | $\frac{2^{-3+p}\sqrt{\pi}\, z^{-1-p}\, p(p-2)}{\Gamma(\frac{3}{2}-p)}$ | $-\frac{2^{-7/2+p}\sqrt{\pi}\, z^{-1-p}\, p(p-2)^2}{\sqrt{3p(p+1)}\,\Gamma(\frac{3}{2}-p)}$ |
| | $\lvert z\rvert >> 1$ | $(1-\frac{p(p-2)Cos(p\pi+2z)}{4z^2})^{1/2}$ | $\sqrt{\frac{4+8p+13p^2+(4+8p-5p^2)Cos(p\pi+2z)}{12p(1+p)}}$ |
| | Range ($p$=-10/9) | $\lvert C(k)\rvert \cong 1$ | $0.258 \le \lvert C(k)\rvert \le 3.87$ |
| $R_c(k)$ | $z \to 0$ | 1 | $\frac{\sqrt{6p(1+p)}}{2-p} \cong 0.277$ |
| | $\lvert z\rvert >> 1$ | 1 | $\sqrt{\frac{12p(1+p)}{4+8p+13p^2+(4+8p-5p^2)Cos(p\pi+2z)}}$ |
| | Range ($p$=-10/9) | | $0.258 \le R_c(k) \le 3.87$ |
| $\Delta n$ | $z \to 0$ | -2(1+$p$) | -2(1+$p$) |
| | $\lvert z\rvert >> 1$ | 0 | 0 |

Table 1. Summarized results for $|C(k)|$, $R_c(k)$ and $\Delta n$